\documentclass[12pt]{article}
\usepackage{latexsym,epsfig,graphicx,amsmath,amssymb,amscd,undertilde,multirow,chicago,psfrag,paralist}
\usepackage{longtable,float,afterpage}
\usepackage[section]{placeins}
\usepackage{diagbox}
\usepackage{epstopdf}
\usepackage{listings}

\newcommand{\E}{\mathbf{E}}
\newcommand{\Bset}{\mathbf{B}}
\newcommand{\Xset}{\mathbf{X}}
\newcommand{\Sset}{\mathbf{S}}

\newcommand{\pr}{{\rm Pr}}

\newcommand{\TAE}{{\rm TAE}}
\newcommand{\MAE}{{\rm MAE}}
\newcommand{\bin}{{\rm bin}}
\newcommand{\enum}{{\rm enum}}

\newcommand{\Arg}{{\rm Arg}}

\newcommand{\atantwo}{{\rm atan2}}
\newcommand{\ivec}{{\boldsymbol{i}}}

\newcommand{\rvec}{\boldsymbol{r}}

\begin{document}
\title{An Algorithm for Computing the Distribution Function of the Generalized Poisson-Binomial Distribution}
\author{Man Zhang and Yili Hong\\
Department of Statistics\\
Virginia Tech\\
Blacksburg, VA 24061, USA
\and
Narayanaswamy Balakrishnan\\
Department of Mathematics and Statistics\\
McMaster University\\
Hamilton, ON, L8S 4K1, Canada
}

\date{\today}

\maketitle

\begin{abstract}
The Poisson-binomial distribution is useful in many applied problems in engineering, actuarial science, and data mining. The Poisson-binomial distribution models the distribution of the sum of independent but not identically distributed Bernoulli random variables whose success probabilities vary. In this paper, we extend the Poisson-binomial distribution to the generalized Poisson-binomial (GPB) distribution. The GPB distribution is defined in cases where the Bernoulli variables can take any two arbitrary values instead of 0 and~1. The GPB distribution is useful in many areas such as voting theory, actuarial science, warranty prediction, and probability theory. With few previous works studying the GPB distribution, we derive the probability distribution via the discrete Fourier transform of the characteristic function of the distribution. We develop an efficient algorithm for computing the distribution function, which uses the fast Fourier transform. We test the accuracy of the developed algorithm upon comparing it with enumeration-based exact method and the results from the binomial distribution. We also study the computational time of the algorithm in various parameter settings. Finally, we discus the factors affecting the computational efficiency of this algorithm, and illustrate the use of the software package.


\end{abstract}

\newpage

\section{Introduction}\label{sec:introduction}
The binomial distribution is defined as the sum of independent and identically distributed random indicators. The cumulative distribution function (cdf) and probability mass function (pmf) of the binomial distribution can be easily computed from closed-form expressions. When those indicators are not identically distributed (i.e., the success probabilities are no longer the same), the sum of those independent and non-identically distributed random indicators is distributed as the Poisson binomial distribution~(e.g., \citeNP{Hong2013}). The Poisson binomial distribution can be considered as a generalization of the binomial distribution. \citeN{Hong2013} developed an algorithm to compute the cdf of the Poisson binomial distribution.

Even though the Poisson-binomial distribution is applicable in a wide range of areas such as in engineering, actuarial science, and data mining, a more general case is often of interest in practice. That is a random indicator may take any two arbitrary values instead of 0 and~1 as defined in the Poisson-binomial distribution. To motivate the research problem, we provide the following examples.

\begin{inparaitem}
\item In voting theory (e.g., \citeNP{Alger2006}), each voter may vote in favor of a proposal with different probabilities. In some settings, some voters may be in a more powerful position, for example, a voter may have two votes. The interest here is in the total number of votes that is in favor of the proposal. The general question is what is the distribution of the total number of votes in favor of the proposal.

\item In warranty cost prediction (e.g., \citeNP{HongMeeker2013}), suppose there is a batch of units in the field and their failure probabilities within one year are different from unit to unit. If one unit fails, the company needs to pay a certain amount of expenses and these expenses will be different from unit to unit. The total expense is of interest in this case, and we are naturally interested in the distribution of the total expenses.

\item In actuarial science (e.g., \citeNP{Pitacco2007}), the amount of insurance payout is related to the payout to each customer and the payout probabilities. The distribution of the total amount of payout is generally of interest.

\item In probability theory, the Rademacher distribution is defined as the distribution of a random variable that with 50\% chance of being either +1 or -1 (\citeNP{MontgomerySmith1990}). The Rademacher series is defined as a weighted sum of a series of independent Rademacher random variables (\citeNP{ChengMCN2014}). The distribution of the Rademacher series is of interest (e.g., \citeNP{DilworthMontgomery-Smith1993}).

\end{inparaitem}

To better describe the problem, we need some notation. Let a random indicator $I_k$ follow the Bernoulli distribution. That is,
$$I_k\sim\textrm{Bernoulli}(p_k), \quad k=1,\dots,n,$$
where $n$ is the total number of indicators. These $I_k$'s are independent but not identically distributed because the $p_k$'s are not necessarily the same. When $I_k=1$, the corresponding value is $b_k$ and when $I_k=0$, the corresponding value is $a_k$, where $a_k<b_k, k=1, \cdots, n$. The random variable of interest is the sum that is given by
\begin{align}\label{eqn:definition}
X=\sum_{k=1}^n a_k(1-I_k)+b_kI_k.
\end{align}
We call $X$ a generalized Poisson-binomial (GPB) random variable. Relating to the voting example, $X$ corresponds to the total number of votes that is in favor of the proposal. When all $a_k=0$ and all $b_k=1$, $X$ reduces to $X=\sum_{k=1}^{n}I_k$, which is the Poisson-binomial random variable. Obviously, when all $p_k$'s are the same, $X$ further reduces to the binomial distribution.

The cdf of $X$ is defined as
$$F(x)=\pr(X\leq x).$$
The computation of the cdf, however, is non-trivial. There is no existing algorithm that can efficiently compute the cdf, despite the usefulness of the distribution in practice. This paper focuses on the development of an efficient algorithm for computing the cdf $F(x)$.

The rest of the paper is organized as follows. Section~\ref{sec:algorithm.dev} discusses the development and implementation of the algorithm. Section~\ref{sec:implementation} covers the validation of the proposed algorithm. Section~\ref{sec:illustration} provides illustration of the algorithm. Finally, Section~\ref{sec:conclusion} presents some concluding remarks.

\section{Computation of Distribution Function}\label{sec:algorithm.dev}
\subsection{Theoretical Formula}\label{sec:enum.cdf}
Because each indicator $I_k$ can take values in $\{0, 1\}$, all possible combinations can be denoted by
\begin{align*}
\Bset=\underbrace{\{0,1\}\times\cdots\times\{0,1\}\times\cdots \times \{0,1\}}_{n \textrm{ times}}.
\end{align*}
Let $\rvec=(r_1, \cdots, r_k, \cdots, r_n)'$ be an element in $\Bset$. The number of elements in $\Bset$ is $2^{n}$. The value of each element of $\rvec$ is either 0 or 1. The set of all possible values of $X$ is
\begin{align*}
\Xset=\left\{x: \textrm{ there exists at least one } \rvec \textrm{ such that } x=\sum_{k=1}^na_kr_k+b_kr_k\right\}.
\end{align*}
Let $a=\sum_{k=1}^{n}a_k$ and $b=\sum_{k=1}^{n}b_k$. The possible values for $X$ ranges from $a$ to $b$, and evidently
\begin{align*}
\Pr(X=x)=\sum_{\rvec\in \Sset_x}\prod_{k=1}^np_k^{r_k}(1-p_k)^{1-r_k},
\end{align*}
where $\Sset_x=\{\rvec: x=\sum_{k=1}^na_kr_k+b_kr_k\}$.

\subsection{Proposed Method Formula Based on Permutation}
An alternative way of calculating the pmf is based on permutation. Let $\{k_1,  k_2, \cdots, k_n\}$ be a permutation of indices $\{1, 2, \cdots, n\}$. A possible value of $X$ is $x=\sum_{l=1}^R a_{k_l}+\sum_{l=R+1}^n b_{k_l}$, for $R=0, 1, \dotsc n$. It is evident that $X=\sum_{l=1}^R a_{k_l}+\sum_{l=R+1}^n b_{k_l}$ if and only if $I_{k_1}=I_{k_2}=\cdots=I_{k_R}=0$ and all other $I_i$'s are 1. The corresponding probability is
$$\Pr(I_{k_1}=I_{k_2}=\cdots=I_{k_R}=0, \textrm{ all other } I_i\textrm{'s}=1)=\left[\prod_{l=1}^R(1-p_{k_l})\right]\times\left[\prod_{l=R+1}^np_{k_l}\right].$$
Thus,
$$\Pr\left(X=\sum_{l=1}^R a_{k_l}+\sum_{l=R+1}^n b_{k_l}\right)=\frac{1}{R!(n-R)!} \sum_{S} \left[\prod_{l=1}^R(1-p_{k_l})\right]\times\left[\prod_{l=R+1}^np_{k_l}\right],$$
where $S$ is the sum over all $n!$ permutations of $\{1, 2, \cdots, n\}$. The pmf of $X$ can be obtained by aggregating those probabilities for those with the same $x$ values.

Note that both the enumeration and permutation based methods can be computationally infeasible for large $n$ (e.g., when $n>30$). When $n=30$, $2^{30}$ is around 1 billion, and $30!$ is around $2.6\times10^{32}$. Thus, a computationally efficient method needs to be developed, which will be presented in next section.

\subsection{Proposed Method}
For the development of the algorithm, we restrict $a_k$ and $b_k$ to be integers. We will discuss non-integer cases later in Section~\ref{sec:illustration}. Let
$$\xi_j=\Pr(X=j+a), j=0,\cdots, m,$$
be the probability mass function (pmf) of $X$, where $a=\sum_{k=1}^{n}a_k$, $b=\sum_{k=1}^{n}b_k$, and $m=b-a$. The objective is to compute the pmf $\{\xi_0, \xi_1,\cdots, \xi_m\}$. The characteristic function (cf) of $X$ is
\begin{align}\label{eqn.cf1}
\varphi(t)=\E[\exp(\ivec tX)]=\sum_{j=0}^{m}\xi_{j}\exp[\ivec t(j+a)],
\end{align}
where $\ivec=\sqrt{-1}$. Alternatively, the cf can be computed as
\begin{align}\label{eqn:cf2}
\varphi(t)&=\E\left\{\exp\left[\ivec t\sum_{k=1}^n a_k(1-I_k)+b_kI_k\right]\right\}\\\nonumber
&=\prod_{k=1}^n\left[(1-p_k)\exp(\ivec ta_k)+p_k\exp(\ivec tb_k)\right].
\end{align}
Linking \eqref{eqn.cf1} and \eqref{eqn:cf2}, we obtain,
\begin{align}\label{eqn:cf.two.sides}
\sum_{j=0}^{m}\xi_{j}\exp(\ivec tj)=\exp[-\ivec ta]\prod_{k=1}^n\left[(1-p_k)\exp(\ivec ta_k)+p_k\exp(\ivec tb_k)\right].
\end{align}
Let $\omega=2\pi/(m+1)$. By substituting $t=\omega l, l=0,1,\cdots,m$ into \eqref{eqn:cf.two.sides}, we obtain
\begin{align}\label{eqn:cf.eqns}
\frac{1}{m+1}\sum_{j=0}^m\xi_j\exp(\ivec\omega lj)&=\frac{1}{m+1}\exp[-\ivec \omega la]\prod_{k=1}^n\left[(1-p_k)\exp(\ivec \omega la_k)+p_k\exp(\ivec \omega lb_k)\right]\\\nonumber
&=\frac{1}{m+1}x_l, \quad l=0,1,\cdots,m.
\end{align}
Here, $$x_l=\exp[-\ivec \omega la]\prod_{k=1}^n\left[(1-p_k)\exp(\ivec \omega la_k)+p_k\exp(\ivec \omega lb_k)\right].$$
The left hand side of \eqref{eqn:cf.eqns} is exactly the inverse discrete Fourier transform (IDFT) of $\{\xi_0,\xi_1,$ $\cdots,\xi_m\}$ (e.g., see \citeNP{Bracewell2000}). Thus, we can recover $\{\xi_0,\xi_1,\cdots,\xi_m\}$ by applying the discrete Fourier transform (DFT) to both sides of \eqref{eqn:cf.eqns}. That is,
\begin{align*}
\xi_j&=\frac{1}{m+1}\sum_{l=0}^{m}\exp(-\ivec\omega lk)x_l.
\end{align*}
One first needs to compute $x_l$ and then one can obtain $\xi_j, j=0,1,\dots,m$. Note that $x_l$ can be represented as $x_l=u_l+\ivec v_l, l=0,1,\dots,m$, where $u_l$ and $v_l$ are the real and imaginary parts of $x_l$, respectively. According to \eqref{eqn:cf.eqns}, $$x_l=\sum_{j=0}^m\xi_j\exp(\ivec\omega lj), l=0,1,\cdots, m.$$
It is evident that $x_0=\sum_{k=0}^n\xi_k=1$. In addition, note that all $\xi_j$'s are real numbers and $\exp[\ivec\omega(m+1)j]=1$. Thus, the conjugate of $x_l$ is
\begin{align*}
\overline{x_l}&=u_l-\ivec v_l=\sum_{j=0}^m\xi_j\exp(-\ivec\omega lj)=\sum_{j=0}^m\xi_j\exp[\ivec\omega(m+1-l)j]\\
&=x_{m+1-l}=u_{m+1-l}+\ivec v_{m+1-l}, \quad l=1,\dots,m.
\end{align*}
We obtain $u_l=u_{m+1-l}$, and $v_l=-v_{m+1-l}$ for $l=1,\dots,m$. Let
$z_{0l}=\cos(-\omega la)+\ivec\sin(-\omega la),$
and
$$z_{kl}=[(1-p_k)\cos(\omega la_k)+p_k\cos(\omega lb_k)]+\ivec[(1-p_k)\sin(\omega la_k)+p_k\sin(\omega lb_k)].$$
We denote $|z_{kl}|$ as the modulus of $z_{kl}$, and $\Arg(z_{kl})$ as the principal value of the argument of $z_{kl}$. We obtain
\begin{align*}
x_l&= \exp\left[\sum_{k=0}^n\log(z_{kl})\right]=\exp\left(\sum_{k=0}^n\log\left\{|z_{kl}|\exp[\ivec \Arg(z_{kl})]\right\}\right)\\
&=\exp\left[\sum_{k=0}^n\log\left(\,|z_{kl}|\,\right)\right]\exp\left[\ivec\sum_{k=0}^n\Arg(z_{kl})\right]\\
&=\exp\left[\sum_{k=0}^n\log\left(\,|z_{kl}|\,\right)\right]\left\{\cos\left[\sum_{k=0}^n\Arg(z_{kl})\right]
+\ivec\sin\left[\sum_{k=0}^n\Arg(z_{kl})\right]\right\}.
\end{align*}
In this case, $|z_{0l}|=1$, $\Arg[z_{0l}]=\atantwo\left\{\sin(-\omega la), \cos(-\omega la)\right\}$, and
\begin{align*}
|z_{kl}|&=\big\{[(1-p_k)\cos(\omega la_k)+p_k\cos(\omega lb_k)]^2+[(1-p_k)\sin(\omega la_k)+p_k\sin(\omega lb_k)]^2\big\}^{\frac{1}{2}},\\
\Arg[z_{kl}]&=\atantwo\left\{[(1-p_k)\sin(\omega la_k)+p_k\sin(\omega lb_k)], [(1-p_k)\cos(\omega la_k)+p_k\cos(\omega lb_k)]\right\},
\end{align*}
where $\atantwo (y,x)$ is defined as
$$\atantwo(y, x) = \begin{cases}
\arctan(\frac y x) & \qquad x > 0 \\
\pi + \arctan(\frac y x) & \qquad y \ge 0 , x < 0 \\
-\pi + \arctan(\frac y x) & \qquad y < 0 , x < 0 \\
\frac{\pi}{2} & \qquad y > 0 , x = 0 \\
-\frac{\pi}{2} & \qquad y < 0 , x = 0 \\
0 & \qquad y = 0, x = 0
\end{cases}.$$
We obtain explicit expressions for $u_l$ and $v_l$ as
\begin{align}\label{eqn:al.bl}
u_l=z_l\cos\left[\sum_{k=0}^n\Arg(z_{kl})\right]\text{ and }v_l=z_l\sin\left[\sum_{k=0}^n\Arg(z_{kl})\right],
\end{align}
where $z_l=\exp\left[\sum_{k=0}^n\log\left(\,|z_{kl}|\,\right)\right], l=1,\dots,m.$

\subsection{The DFT-CF Algorithm and Implementation}
The following algorithm is used to compute the pmf $\xi_j$, for $j=0,1,\cdots,m.$ \\[.8em]
\textbf{The DFT-CF Algorithm:}
\begin{enumerate}
\item We first assign $x_0=1$. Then, we compute the real and imaginary parts of $x_l$ by using the formulae in \eqref{eqn:al.bl}, $l=1,\dots [m/2]$,  and $[\,\cdot\,]$ is the ceiling function;
\item We compute the real and imaginary parts of $x_l$ by using the formula $u_l=u_{m+1-l}$, and $v_l=-v_{m+1-l}$, $l=[m/2]+1,\dots, m$.
\item We then apply the fast Fourier transform (FFT) algorithm to the set $\{x_0/(m+1), x_1/(m+1),\dots,x_n/(m+1)\}$ to obtain $\{\xi_0,\xi_1,\dots, \xi_m\}$.
\end{enumerate}

The DFT-CF algorithm has been implemented in C and it can be called from R. We also wrap the major functions into an R package \texttt{GPB} (\citeNP{RGPB}). The use of the R package will be illustrated in Section~\ref{sec:illustration}.

\section{Algorithm Validation}\label{sec:implementation}
This section focuses on the validation of the developed algorithm.

\subsection{Accuracy Comparison with an Exact Method}
We develop an enumeration-based algorithm to compute the exact cdf of the GPB distribution based on theoretical formula. Then, we use the maximum absolute error (MAE) and the total absolute error (TAE) as accuracy metrics by comparing cdf calculated with the DFT-CF algorithm and the enumeration-based method for different values of $n, a_k, b_k$ and $p_k$. The maximum absolute error (MAE) is defined as
$$\MAE=\max_{x}| F(x)- F_{\enum} (x)|,$$
while the total absolute error (TAE) is defined as
$$\TAE=\sum_{x=a}^{b} |F(x)- F_{\enum} (x)|,$$
where $ F(x)$ is the cdf computed by using the DFT-CF algorithm and $ F_{\enum}(x)$ is the cdf computed by using the enumeration formula in Section~\ref{sec:enum.cdf}.

The accuracy test results are shown in Table~\ref{table:accuracy1} for different parameter settings for $n, a_k, b_k$ and $p$. All computations were done on Linux 64-bit server with Intel Xeon CPU (E5-2680, 2.50GHz) and 263 GB RAM. Due to the complex enumeration calculation, the exact method can only handle less than 30 random indicators (i.e., $n=30$) under the capacity of the computer server. Table~\ref{tab:hresult} shows the accuracy of the cdf calculated with the DFT-CF algorithm for various values of $n, a_k, b_k$ and $p$. The MAE are generally less than $5\times10^{-15}$ and the TAE are less than $5\times10^{-14}$ for the DFT-CF algorithm, when $n$ is less than 20. Overall, the results show that the DFT-CF algorithm can accurately compute the cdf of the GPB distribution.

\begin{table}
\caption{Accuracy of the DFT-CF algorithm compared with the enumeration method.}\label{tab:hresult}
\vspace{1ex}
\centering
\begin{tabular}{c c c c c c c}
\hline\hline
$n$& $a$ & $b$ & $\min(p_k)$ & $\max(p_k)$ & MAE & TAE \\\hline
10  &10    &20 &0.01&0.50        &$6.7\times10^{-16}$&$4.4\times10^{-14}$ \\
10  &10    &20 &0.50&0.99        &$7.3\times10^{-16}$&$2.8\times10^{-15}$ \\
10  &10    &20 &0.01&0.99        &$9.4\times10^{-16}$&$3.8\times10^{-15}$ \\\hline
10  &10    &50 &0.01&0.50        &$6.7\times10^{-16}$&$4.4\times10^{-14}$ \\
10  &10    &50 &0.50&0.99        &$7.3\times10^{-16}$&$2.8\times10^{-15}$ \\
10  &10    &50 &0.01&0.99        &$9.4\times10^{-16}$&$3.8\times10^{-15}$ \\\hline
10  &50    &100&0.01&0.50        &$6.7\times10^{-16}$&$4.4\times10^{-14}$ \\
10  &50    &100&0.50&0.99        &$7.3\times10^{-16}$&$4.6\times10^{-15}$ \\
10  &50    &100&0.01&0.99        &$9.4\times10^{-16}$&$4.5\times10^{-15}$ \\\hline
20  &20    &40 &0.01&0.50        &$4.4\times10^{-16}$&$4.1\times10^{-15}$ \\
20  &20    &40 &0.50&0.99        &$6.7\times10^{-16}$&$4.1\times10^{-14}$ \\
20  &20    &40 &0.01&0.99        &$1.3\times10^{-15}$&$1.1\times10^{-14}$ \\\hline
20  &40    &100&0.01&0.50        &$4.4\times10^{-16}$&$4.1\times10^{-15}$ \\
20  &40    &100&0.50&0.99        &$6.7\times10^{-16}$&$4.1\times10^{-14}$ \\
20  &40    &100&0.01&0.99        &$1.3\times10^{-15}$&$1.1\times10^{-14}$ \\\hline
20  &100   &200&0.01&0.50        &$4.4\times10^{-16}$&$4.1\times10^{-15}$ \\
20  &100   &200&0.50&0.99        &$6.7\times10^{-16}$&$4.1\times10^{-14}$ \\
20  &100   &200&0.01&0.99        &$1.3\times10^{-15}$&$1.1\times10^{-14}$ \\
\hline\hline
\end{tabular}
\label{table:accuracy1}
\end{table}

\subsection{Accuracy Comparison with the Binomial Distribution}
To test the accuracy of the DFT-CF algorithm for large values of $m$ and $n$, we compare the cdf computed by the DFT-CF algorithm with that of binomial distributions. As mentioned earlier, the binomial distribution is a special case of the GPB distribution when all $p_k$'s are the same, and $a_k=0, b_k=1$ for all $n$ random indicators. Thus in this comparison setting, we let $p_k=p$ to be the same, and $ a_k=0, b_k=1$. That is,
$$X = \sum_{k=1}^n a_k(1-I_k) + b_k I_k= \sum_{k=1} ^n I_k.$$
The exact pmf of X can be calculated from the binomial distribution as
$$\Pr(X=x)= {{n}\choose{x}} p^ x (1-p)^ {n-x}.$$
Here again, the MAE and TAE are used as accuracy metrics, which are given by
$$\MAE=\max_{x}| F(k)- F_{\bin} (k)|,\quad \textrm{ and } \quad \TAE=\sum_{x=0}^n | F(x)- F_{\bin} (x)|,$$
where $ F(x)$ is the cdf computed by the DFT-CF algorithm and $ F_{\bin}(x)$ is the cdf computed using the binomial distribution function implemented in R~\citeyear{R}. The accuracy test results are shown in Table~\ref{table:accuracy2} with different parameter settings for $n, a_k, b_k$ and $p$. Basically, the TAE and MAE accumulate as $n, p$ and $a_k, b_k$ increases. When $n$ is less than 10,000, the MAE is within $1\times10^{-12}$ and TAE is under $1\times10^{-8}$. The results in Table~\ref{table:accuracy2} show that the DFT-CF algorithm can accurately compute the cdf for large $n$. With no available efficient algorithm developed for computing the GPB distribution, the proposed method provides an important alternative to model real-life applications.

\begin{table}
\caption{Accuracy of the DFT-CF algorithm compared with the binomial distribution, with $a=0$, and $b=n$.}\label{table:accuracy2}
\vspace{1ex}
\centering
\begin{tabular}{cccc|cccc}\hline\hline
$n$ &$p$ &MAE 			     & TAE &           $n$  &$p$    & MAE                       & TAE \\\hline
10   &0.01  &$8.9\times10^{-16}$&$3.9\times10^{-15}$&2{,}000  &0.01   &$2.9\times10^{-14}$&$2.3\times10^{-11}$\\
10   &0.50  &$4.4\times10^{-16}$&$1.6\times10^{-15}$&2{,}000  &0.50   &$1.4\times10^{-13}$&$1.1\times10^{-10}$\\
10   &0.90  &$6.7\times10^{-16}$&$3.2\times10^{-15}$&2{,}000  &0.90   &$4.2\times10^{-13}$&$3.4\times10^{-10}$\\
20   &0.01  &$4.4\times10^{-16}$&$4.7\times10^{-15}$&5{,}000  &0.01   &$1.3\times10^{-13}$&$3.2\times10^{-10}$\\
20   &0.50  &$7.1\times10^{-16}$&$6.3\times10^{-15}$&5{,}000  &0.50   &$4.3\times10^{-13}$&$6.4\times10^{-10}$\\
20   &0.90  &$1.9\times10^{-15}$&$1.8\times10^{-14}$&5{,}000  &0.90   &$7.1\times10^{-13}$&$1.1\times10^{-9}$\\
50   &0.01  &$2.0\times10^{-15}$&$5.3\times10^{-14}$&10{,}000 &0.01   &$4.2\times10^{-13}$&$1.8\times10^{-9}$\\
50   &0.50  &$2.9\times10^{-15}$&$5.2\times10^{-14}$&10{,}000 &0.50   &$1.1\times10^{-12}$&$3.2\times10^{-9}$\\
50   &0.90  &$3.5\times10^{-15}$&$5.3\times10^{-14}$&10{,}000 &0.90   &$1.6\times10^{-12}$&$4.6\times10^{-9}$\\
100  &0.01  &$1.4\times10^{-14}$&$5.7\times10^{-13}$&20{,}000 &0.01   &$6.9\times10^{-13}$&$3.6\times10^{-9}$\\
100  &0.50  &$1.9\times10^{-15}$&$3.7\times10^{-14}$&20{,}000 &0.50   &$2.7\times10^{-12}$&$1.7\times10^{-8}$\\
100  &0.90  &$7.4\times10^{-15}$&$2.7\times10^{-13}$&20{,}000 &0.90   &$5.4\times10^{-12}$&$3.9\times10^{-8}$\\
200  &0.01  &$7.8\times10^{-15}$&$8.8\times10^{-13}$&50{,}000 &0.01   &$3.1\times10^{-12}$&$6.7\times10^{-8}$\\
200  &0.50  &$5.3\times10^{-15}$&$4.9\times10^{-13}$&50{,}000 &0.50   &$9.8\times10^{-12}$&$1.3\times10^{-7}$\\
200  &0.90  &$3.0\times10^{-14}$&$1.9\times10^{-12}$&50{,}000 &0.90   &$1.6\times10^{-11}$&$1.8\times10^{-7}$\\
500  &0.01  &$2.4\times10^{-14}$&$5.0\times10^{-12}$&100{,}000&0.01   &$5.7\times10^{-12}$&$2.0\times10^{-7}$\\
500  &0.50  &$2.4\times10^{-14}$&$5.5\times10^{-12}$&100{,}000&0.50   &$1.6\times10^{-11}$&$3.4\times10^{-7}$\\
500  &0.90  &$6.7\times10^{-14}$&$1.7\times10^{-11}$&100{,}000&0.90   &$4.3\times10^{-11}$&$1.2\times10^{-6}$\\
1{,}000 &0.01  &$5.2\times10^{-14}$&$2.1\times10^{-11}$\\
1{,}000 &0.50  &$5.8\times10^{-14}$&$2.0\times10^{-11}$\\
1{,}000 &0.90  &$1.8\times10^{-13}$&$7.5\times10^{-11}$\\\hline\hline
\end{tabular}
\end{table}

\subsection{Computational Efficiency}
The computational time of the DFT-CF algorithm is mostly determined by $n$ and $m$. Note that $m=b-a=\sum_{k=1}^n b_k-\sum_{k=1}^n a_k$. We first consider the computational time when $n$ is large. We first choose 10 $p$'s from [0.01, 0.99], and for different values of $n$ and $m$, we set $p_k=p, a_k=0, b_k=1$ for all $n$ indicators. The time for calculating the entire cdf using the DFT-CF algorithm is averaged across 10 $p$'s, which are shown in Table~\ref{table:time}. The unit of computation time is second.

\begin{figure}
\centering
\includegraphics[width=0.6\textwidth]{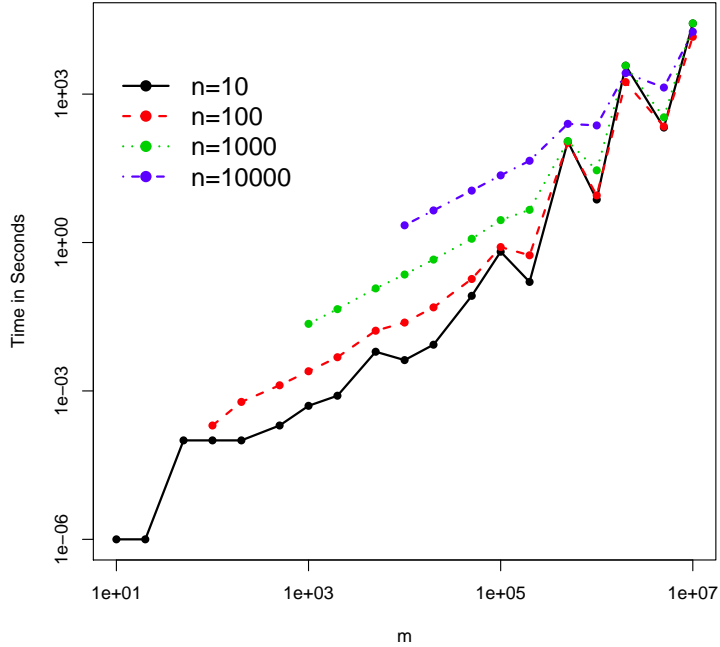}
\caption{Average computational time of the DFT-CF algorithm for different choices of $n$ and $m$. The x-axis and y-axis are on the log-scale.}
\label{figure:time}
\end{figure}

Figure~\ref{figure:time} visualizes the results in Table~\ref{table:time}. Both the x-axis and y-axis are on log scales. Each line indicates the average computation time for a specific $n$, where $n=10$, 100, 1{,}000, and 10,000. Figure~\ref{figure:time} shows that as $n$ increases, the computational time increases exponentially. The computational time is negligible (less than 10 milliseconds) when $n \leq 100$. When $n$ is fixed at 10, 100, 1{,}000, and 10{,}000, the general average computational time increases as $m$ increases. However, there are minor drops in computational time even when $m$ increases to big numbers. The DFT-CF algorithm can work out $n \leq 10{,}000$ and $m \leq 1{,}000{,}000$ within 5 minutes. As $n$ exceeds 10{,}000 and $m$ exceeds 1{,}000{,}000, the DFT-CF algorithm requires more than 5 minutes. Overall, the DFT-CF algorithm shows reasonable computational efficiency.

\begin{table}
\caption{Average computation time for the DFT-CF algorithm over $p$ for different choices of $n$ and $m$.} \label{tab:hresult}
\vspace{1ex}
\centering
\begin{tabular}{r|rrrr}
\hline
\diagbox{$m$}{$n$} & 10 & 100 & 1{,}000 & 10{,}000 \\
\hline
10&0.000 \\
20&0.000\\
50&0.000\\
100&0.000&0.000\\
200&0.000&0.001\\
500&0.000&0.001\\
1{,}000&0.001&0.003&0.023\\
2{,}000&0.001&0.005&0.045\\
5{,}000&0.006&0.017&0.118\\
10,000&0.004&0.024&0.225&2.228\\
20,000&0.009&0.049&0.452&4.456\\
50,000&0.084&0.183&1.191&11.207\\
100,000&0.652&0.809&2.837&22.786\\
200,000&0.160&0.552&4.579&44.933\\
500,000&110.423&103.886&112.395&250.482\\
1,000,000&7.406&8.973&28.826&232.975\\
2,000,000&3765.094&1751.390&3753.482&2657.690\\
5,000,000&211.050&221.361&339.458&1358.968\\
10,000,000&26850.920&14401.410&26927.20&18215.330\\
\hline
\end{tabular}
\label{table:time}
\end{table}

\section{Illustrations}\label{sec:illustration}
\subsection{The Software Package}
The DFT-CF algorithm has been implemented in the R package \texttt{GPB}, \citeN{RGPB}, which can be downloaded from the Comprehensive R Archive Network (http://cran.r-project.org/). The R functions for computing the cdf, pmf, quantile function and random number generation are all available in the R package. For example, the function \texttt{pgpb()} takes $p_k$'s, $a_k$'s and $b_k$'s as input and computes the cdf of the distribution. See the following R code for reference.
\begin{verbatim}
library(GPB)
pgpb(kk=6:9,pp=c(0.1,0.2,0.3),aval=c(1,2,3),bval=c(2,3,4),
wts=c(1,1,1))
\end{verbatim}
Here, \texttt{kk} is $x$ (i.e., the values where the cdf to be evaluated), \texttt{pp} is the vector of $p_k$'s, \texttt{aval} is the vector of $a_k$'s, \texttt{bval} is the vector of $b_k$'s, and \texttt{wts} is the vector of weights for $p_k$'s.

\subsection{Computational Tips}
There are a few tips in the use of the algorithm.  One can use the weights option to speed up the computing when the $p_k$'s are the same. For example, if there are 1{,}000 random indicators with same $p$, instead of replicating the same random indicator 1{,}000 times, one can specify \texttt{wts}=1{,}000. The use of \texttt{wts} argument will speed up the computing due to the implementation of the algorithm.

The proposed algorithm can be slow in cases where $n$ and $m$ are extremely large. In some cases, the problem can be eased by taking out the greatest common divider of $a_k$'s and $b_k$'s. For example, when $k=1, 2, \dots, 10$, $a_k = {10, 20, \dots, 100}$, and $b_k = {100, 200, \dots, 1{,}000}$, the cdf is equivalent to the cdf when $a_k = {1, 2, \dots, 10}$ and $b_k = {10, 20, \dots, 100}$, with a multiplier of 10 for the support values. Thus, by taking out the great common divider, we can shorten the computational time especially when $m$ is large and the common dividers between $a_k$'s and $b_k$'s are large. See the following R code for reference.
\begin{verbatim}
pgpb(kk=seq(10,100,by=10), pp=c(.1, .2, .3), aval=c(10,20,30),
bval=c(20,30,40), wts=c(1,1,1))
pgpb(kk=1:10, pp=c(.1, .2, .3), aval=c(1,2,3), bval=c(2,3,4),
wts=c(1,1,1))
\end{verbatim}

Though the algorithm is derived based on $a_k$ and $b_k$ being integers, it can be applied to non-integer cases by multiplying powers of $10$ to convert decimal digits into integers. For example, if $a_k =\{0.5, 1.5, \dots, 9.5\}$, $b_k = \{1, 2, \dots, 10\}$, and cdf needs to be computed at $x=50.5$, then we can multiply the set of $a_k$, $b_k$ and $x$ by 10. The cdf value is the same as being computed at $x=505$ with $a_k = \{5, 15, \dots, 95\}$ and $b_k = \{10, 20, \dots, 100\}$. Note that the multiplication by powers of 10 increases $m$ and correspondingly increases the computation time. However, this process can extend the proposed algorithm to the cases of non-integer numbers. See the following R code for reference.
\begin{verbatim}
aval=seq(0.5,9.5,by=1)*10
bval=seq(1,10,by=1)*10
pgpb(kk=50.5*10,pp=seq(0.1,0.5,length.out=10),aval=aval,
bval=bval,wts=rep(1,10))
\end{verbatim}

\section{Concluding Remarks}\label{sec:conclusion}
In this paper, we consider the GPB distribution, which has applications in many areas. We derive a closed-form expression for the cdf by using the DFT-CF algorithm. We demonstrate that the proposed algorithm is accurate in terms of error as compared to an enumeration-based exact method and the binomial distribution. We further show the computational efficiency and the limitation of the DFT-CF algorithm in numerical analysis for different settings of $n$ and $m$. The DFT-CF algorithm is generally accurate (with the TAE under $1\times10^{-8}$) and computationally efficient (less than five minutes for computing) when $n$ is less than 10{,}000.  The DFT-CF algorithm can be extended to non-integer numbers as well. The proposed method has been implemented in an R package named \texttt{GPB}.

In Section~\ref{sec:introduction}, we discus several possible areas of applications for the GPB distribution. However, the application areas of the GPB distribution is much broader. For example, it can be useful areas such as econometrics (e.g., \citeNP{DuffieSaitaWang2007}), data mining (e.g., \citeNP{TangPeterson2011}), bioinformatics (e.g., \shortciteNP{Niidaetal2012}), renewable energy (\citeNP{BossavyGirardKariniotakis2012}), and survey sampling (e.g., \citeNP{ChenLiu1997}), in which cases there are costs associated with random indicators with different success probabilities. The implementation of the developed algorithm in R makes it convenient for the practitioners.

\end{document}